\documentclass[preprint,showpacs,preprintnumbers,amsmath,amssymb]{revtex4}

\usepackage{graphicx,color}

\begin{document}

\title{
Deflection angle of light for an observer and source 
at finite distance from a rotating wormhole} 
\author{Toshiaki Ono}
\author{Asahi Ishihara}
\author{Hideki Asada} 
%\email{asada@phys.hirosaki-u.ac.jp}
\affiliation{
Graduate School of Science and Technology, Hirosaki University,
Aomori 036-8561, Japan} 
\date{\today}

\begin{abstract} 
By using a method improved with a generalized optical metric, 
the deflection of light 
for an observer and source at finite distance from a lens object 
in a stationary, axisymmetric and asymptotically flat spacetime 
has been recently discussed 
[Ono, Ishihara, Asada, Phys. Rev. D {\bf 96}, 104037 (2017)]. 
By using this method, in the weak field approximation, 
we study the deflection angle of light 
for an observer and source at finite distance from a rotating Teo wormhole, 
especially by taking account of 
the contribution from the geodesic curvature of the light ray 
in a space associated with the generalized optical metric. 
Our result of the deflection angle of light 
is compared with a recent work 
on the same wormhole but limited within the asymptotic source and observer 
[Jusufi, \"Ovg\"un, Phys. Rev. D {\bf 97}, 024042, (2018)], 
in which they employ another approach proposed by Werner 
with using the Nazim's osculating Riemannian construction method 
via the Randers-Finsler metric. 
We show that the two different methods give the same result 
in the asymptotic limit. 
We obtain also the corrections to the deflection angle 
due to the finite distance from the rotating wormhole. 
\end{abstract}


\maketitle

\section{Introduction}
Studies on wormholes can be dated back to the celebrated paper 
by Einstein and Rosen \cite{ER}, 
in which they investigated what is a particle 
in the theory of general relativity, and 
consequently they noticed a spacetime bridge connecting 
two distinct spacetime events, called Einstein-Rosen bridges. 
Decades later, Wheeler argued that such spacetime bridges should 
be unstable even for a traveling photon \cite{Wheeler}. 
Misner and Wheeler dubbed such a handle of multiply-connected spacetime wormholes 
\cite{MW}. 
Morris, Thorne, and Yurtsever, nevertheless, discussed traversable 
wormholes by holding a throat of the wormholes open with 
hypothetical exotic matter (that must have negative energy 
in the framework of general relativity) 
\cite{MTY}. 
Later, other types of traversable wormholes were found 
as allowable solutions to Einstein equation, especially 
in a 1989 paper by Matt Visser \cite{Visser}, 
in which a spacetime tunnel through the wormhole can be constructed 
where a shortcut path does not pass 
through a region of such exotic matter. 
This type of wormhole models are called thin-shell wormholes. 
See Ref. \cite{Visser-Book} for comprehensive reviews on wormholes. 
In the Gauss-Bonnet gravity (an alternative to the theory of 
general relativity), however, 
exotic matter is not required for wormholes to exist 
\cite{GBWH}. 
The latter wormhole model is based on an idea of modifying 
the left hand side (namely, the geometrical side) of Einstein equation, 
while the former models are due to some modifications 
of the right hand side, especially inclusions of hypothetical exotic matter. 

Null and causal structures of such wormhole spacetimes are 
expected to be very different from those around stellar objects 
and even those in black hole spacetimes. 
Therefore, the light propagation in wormhole spacetimes 
has attracted a lot of interest. 
The deflection of light in Ellis wormhole was first discussed 
by Chetouani and Clement \cite{CC,Clement}. 
The gravitational lensing as an observational probe of wormholes 
was investigated 
\cite{Perlick,Safonova,Shatskii,Nandi,BP,Abe,Toki,Tsukamoto,Kitamura,Sakai,JOB,Asada}. 
In the weak field approximation, 
the deflection angle of light was derived in terms of the inverse power 
of the photon impact parameter, for instance by Dey and Sen 
\cite{DS}. 
However, Nakajima and Asada showed that this result 
breaks down at the next-to-leading order, though 
the leading order term is correct \cite{Nakajima}. 
This problem occurs due to the regularity at the center of wormholes 
and therefore some methods valid for black holes no longer work 
for wormholes. 
On the observational side of wormholes, 
Takahashi and Asada showed that 
the Sloan Digital Sky Survey Quasar Lens Search (SQLS) put 
the upper bound on the cosmic abundance of Ellis wormholes 
\cite{Takahashi}. 

Most of the work on the wormhole lensing mentioned above 
is for non-rotating wormholes. 
Very recently, Jusufi and \"Ovg\"un \cite{JO} discussed the gravitational lensing 
by rotating Teo wormholes 
\cite{Teo}, 
in which they use Gibbons-Werner approach based on the Gauss-Bonnet theorem 
\cite{GW}. 
An extension of the Gibbons-Werner approach for calculating 
the deflection of light for the case of a Kerr black hole 
was done by Werner 
\cite{Werner2012}, 
in which he used Nazim's method of constructing the osculating 
Riemannian manifold and computed the Randers-Finsler form of the metric 
for the Kerr spacetime. 
To be more precise, Jusufi and \"Ovg\"un employed Werner's method 
to calculate the deflection angle of light 
for the asymptotic observer and source 
in the weak field approximation of a rotating Teo wormhole. 
The condition that the observer and source are located 
at the null infinity is a requirement for using Werner's method, 
because the Werner's extension 
by using the Nazim's osculating Riemannian method 
needs that two ends of the light ray 
(corresponding to the observer and source, respectively) 
are in a Euclidean space. 
We should note that it is an open issue how to define angles 
in the Finsler geometry, though angles are well-defined 
in Euclidean regions of the Finsler geometry. 

The main purpose of this paper is to discuss 
the deflection of light for an observer and source 
at finite distance from a rotating Teo wormhole 
as the gravitational lens. 
For this purpose, we shall use a formulation 
developed in Ref. \cite{OIA2017}, 
which we shall call {\it generalized optical metric method} henceforth. 

The method for investigating the light propagation 
in a static and spherically symmetric spacetime  
was reexamined by Gibbons and Werner, 
who discussed 
a problem of how to determine a curve on a spatial surface 
in the optical geometry, 
where the metric used in the optical geometry 
was first called the optical metric \cite{GW}. 
The idea of what Gibbons and Werner call the optical geometry 
may be related with the optical reference geometry 
that was used to describe inertial forces in general relativity 
by Abramowicz et al. 
\cite{Abramowicz}, 
and may be connected also with the idea of the optical 3-geometry 
that was introduced to discuss thermal Green's functions 
for black holes by Gibbons and Perry 
\cite{GP}. 
The optical geometry 
may be also called the optical reference geometry or Fermat
geometry \cite{Werner2012}. 
The merit of the optical metric is that the arc length along 
the light ray with this metric is directly related 
with the time associated with the timelike Killing vector, 
when the spacetime is stationary. 
Namely, the optical metric describes the Fermat's principle 
for the light propagation in a manner simpler than 
other spatial projections of the four-dimensional metric 
such as the intrinsic metric in the ADM formulation.  
The generalized optical method 
is an improved method 
for calculating the deflection angle of light 
especially for the non-asymptotic observer and source 
with the Weyl-Lewis-Papapetrou metric form of 
a stationary, axisymmetric and asymptotically flat spacetime 
(but in the polar coordinates, though it is usually described 
in the cylindrical coordinates \cite{Lewis,LR,Papapetrou}), 
by extending an earlier work 
on static, spherically symmetric 
and asymptotically flat spacetimes 
\cite{Ishihara}. 
The generalized optical metric method 
has been used for discussions on the light deflection 
for the case of Kerr black holes 
\cite{OIA2017}.

There are the pros and cons in the generalized optical metric method. 
The merit of this method 
is that it enables us to calculate the light deflection 
not only for asymptotic observer and source
but also for non-asymptotic cases. 
As stated already, Werner's method, which was used 
by Jusufi and \"Ovg\"un, is currently limited within 
the case of asymptotic observer and source, 
because the observer and source are needed to be 
in a Euclidean space of the Finsler geometry. 
The price for using the generalized optical metric method 
is that we have to take account of the geodesic curvature 
of the light ray in the optical geometry and 
have to do the path integral of the geodesic curvature. 
We note that the light ray is not necessarily 
geodesic in the optical geometry, 
though the light ray follows the null geodesic 
in a four-dimensional spacetime 
\cite{OIA2017}. 
In the present paper, we shall explicitly calculate the geodesic curvature 
in the optical geometry for rotating Teo wormholes 
and perform its path integral. 
A point is that a light ray in Werner's approach 
is treated as a curve in 
a space described by the Randers-Finsler type metric, 
while the generalized optical metric approach discusses a light ray 
as a curve in a space that is defined by introducing the optical metric. 
Two spaces in the two methods are different from each other. 
Therefore, it is important to ask whether both methods 
give the same deflection angle of light, even if the same limiting case 
as the asymptotic observer and source is taken. 
If the deflection angle depended on these calculation methods, 
it might not be useful for gravitational lensing observations. 
We shall show that it is not the case.  
Corrections for the finite distance cases will be also discussed.

In the rest of this paper, the observer is called 
the receiver (R), in order to avoid a confusion in notations 
between the observer and the origin of the coordinates (O). 
This paper is organized as follows. 
Section II describes a rotating Teo wormhole and its optical metric form. 
In Section III, we perform detailed calculations 
of the Gaussian curvature and geodesic curvature 
to obtain the deflection angle of light 
in the weak field approximation of the rotating Teo wormhole. 
A comparison with the earlier work \cite{JO} is also done. 
Section IV is devoted to the conclusion. 
We use the unit of $c=1$ throughout this paper.

\section{Generalized optical metric for rotating Teo wormhole}
\subsection{Rotating Teo wormhole}

A general form of a static axially symmetric rotating wormhole was 
first described by Teo in Ref. \cite{Teo}. 
Its spacetime metric reads 
\begin{align}
ds^2=&-N^2dt^2+\frac{dr^2}{1-\frac{b_0}{r}}
+r^2H^2\Big[d\theta^2+\sin^2\theta(d\phi-\omega dt)^2 \Big] ,
\label{ds2}
\end{align}
where 

the coordinates are 
$-\infty < t < +\infty$, 
$b_0 \leq r < +\infty$, 
$0 \leq \theta \leq \pi$, 
$0 \leq \phi \leq 2\pi$ 
and we denote 
\begin{align}
N=&H=1+\frac{d(4a\cos\theta)^2}{r}, 
\label{N}
\\
\omega=&\frac{2a}{r^3}. 
\label{omega}
\end{align}

The Teo wormhole by Eq. (\ref{ds2}) 
is a rotating generalization of the static Morris-Thorne 
wormhole. 
A rigidly rotating wormhole would be a case of 
$N = H = 1$ and $\omega = const.$ 
The spacetime of Teo is stationary and axially symmetric 
and asymptotically flat, 
and the spatial coordinates $r$, $\theta$ and $\phi$ 
coincide asymptotically with 
the spherical coordinates of a flat space. 
Here, 
$b_0$ denotes the throat radius of the wormhole 
where two identical asymptotically flat regions 
are joined together at the throat $r = b_0$. 
The parameter $a$ is the total angular momentum of the wormhole, 
and the parameter $\omega$ is the angular velocity of the wormhole 
relative to the asymptotic rest frame, 
which gives rise to the Lense-Thirring effect in general relativity. 

As already noticed by Teo \cite{Teo}, 
the wormhole metric in Eq. (\ref{ds2}) 
violates the null energy condition. 
The wormhole (\ref{ds2}) has no singularities 
in the curvature tensor and no event horizon. 
The Teo wormhole metric is a purely geometrical object in 
the sense that the metric does not take account of 
the stress-energy tensor in the Einstein equation. 
As for the possible matter source of a rotating wormhole,  
we refer to \cite{BH}, 
in which general requirements on the stress-energy tensor 
were discussed to generate a uniformly rotating wormhole. 
Here, we are just interested in the geometry of spacetime  (\ref{ds2}) 
as being an exact solution of the gravitational field equations.

\subsection{Optical metric}
Following Ref. \cite{OIA2017}, 
we define the generalized optical metric $\gamma_{ij}$ 
($i, j = 1, 2, 3$)
by a relation as 
\begin{align}
dt=& \sqrt{\gamma_{ij} dx^i dx^j} +\beta_i dx^i , 
\label{opt} 
\end{align}
which is immediately obtained by solving the null condition ($ds^2 = 0$) 
for $dt$. 
Note that $\gamma_{ij}$ is not the induced metric in the ADM formalism. 

For the rotating Teo wormhole by Eq. (\ref{ds2}), 
we find the components of the generalized optical metric 
as 
\begin{align}
\gamma_{ij}dx^idx^j=&
\frac{r^7}{(r-b_0) \left(r^4-4a^2\sin^2\theta \right) 
\left(16da^2\cos^2\theta +r\right)^2}dr^2 \notag\\
&+\frac{r^6}{r^4-4a^2\sin^2\theta } d\theta^2
+\frac{r^{10}\sin^2\theta }{\left(r^4-4a^2\sin^2\theta \right)^2}d\phi^2 . 
\label{gamma}
\end{align}
We obtain the components of $\beta_i$ as 
\begin{align}
\beta_idx^i=&-\frac{2 a r^3 \sin^2\theta }{r^4-4 a^2 \sin^2\theta } d\phi .
\end{align}

In the rest of the paper, we focus on the light rays 
in the equatorial plane, 
namely $\theta = \pi/2$. 
Then, the constant $d$ in the metric does not appear.

\section{Deflection angle of light by a rotating Teo wormhole} 
\subsection{Deflection angle of light}
Let us begin this section with briefly summarizing 
the generalized optical metric method that enables us to 
calculate the deflection angle of light 
for non-asymptotic receiver (denoted as $R$) 
and source (denoted as $S$) \cite{OIA2017}.

We define the deflection angle of light as \cite{OIA2017}
\begin{equation}
\alpha \equiv \Psi_R - \Psi_S + \phi_{RS} . 
\label{alpha-axial}
\end{equation} 
Here, $\Psi_R$ and $\Psi_S$ are angles between the light ray tangent 
and the radial direction from the lens object, 
defined in a covariant manner using the generalized optical metric, 
at the receiver location and the source, respectively. 
On the other hand, 
$\phi_{RS}$ is the coordinate angle between the receiver and source, 
where the coordinate angle is associated with the rotational Killing vector 
in the spacetime. 
If the space under study is Euclidean, this $\alpha$ becomes 
the deflection angle of the curve. 
This is consistent with the thin lens approximation 
in the standard theory of gravitational lensing. 

By using the Gauss-Bonnet theorem \cite{GB-theorem,Math}, 
Eq. (\ref{alpha-axial}) can be recast into \cite{OIA2017}
\begin{align}
\alpha 
=-\iint_{{}^{\infty}_{R}\square^{\infty}_{S}} K dS 
+ \int_{S}^{R} 
\kappa_g d\ell , 
\label{GB-axial}
\end{align} 
where 
$K$ is defined as the Gaussian curvature 
at some point on the two-dimensional surface, 
$dS$ denotes the infinitesimal surface element 
defined with 
$\gamma^{(2)}_{ij}$ where 
$\gamma^{(2)}_{ij}$ denotes the two-dimensional metric 
in the equatorial plane ($\theta = \pi/2$) and reads: 
$\displaystyle{\gamma^{(2)}_{ij}dx^idx^j=
\frac{r^5}{(r-b_0)(r^4-4a^2)}dr^2+\frac{r^{10}}{(r^4-4a^2)^2}d\phi^2}$. 
${}^{\infty}_{R}\square^{\infty}_{S}$ denotes 
a quadrilateral embedded in a curved space with $\gamma_{ij}$, 
$\kappa_g$ denotes the geodesic curvature of the light ray in this space 
and $d\ell$ is an arc length defined with the generalized optical metric 
(See Fig. 2 in Ref. \cite{OIA2017}). 
It is shown by Asada and Kasai that 
this $d\ell$ for the light ray is an affine parameter \cite{AK}. 
Note that only the surface integral term appears in the right hand side 
of Eq. (\ref{GB-axial}) if $\beta_i =0$ 
(See \cite{Ishihara}), 
and 
the path integral term is proportional to the total angular momentum 
of the wormhole (as shown in Subsection IIIC), 
hence caused by rotational (i.e. Lense-Thirring) 
effects of the spacetime. 
We shall make detailed calculations of the R.H.S. of 
Eq. (\ref{GB-axial}) below.

\subsection{Gaussian curvature}
For the equatorial case of a rotating Teo wormhole, 
the Gaussian curvature in the weak field approximation is calculated as 
\begin{align}
K=&\frac{R_{r\phi r\phi}}{\det\gamma^{(2)}_{ij}}
\notag\\
=&\frac{1}{\sqrt{\det\gamma^{(2)}_{ij}}}\Big[\frac{\partial}{\partial\phi}
\Big(\frac{\sqrt{\det\gamma^{(2)}_{ij}}}{\gamma^{(2)}_{rr}}\Gamma^{\phi}_{~rr}\Big)
-\frac{\partial}{\partial r}
\Big(\frac{\sqrt{\det\gamma^{(2)}_{ij}}}{\gamma^{(2)}_{rr}}\Gamma^{\phi}_{~r\phi}\Big)\Big] 
\notag\\
=&-\frac{b_0}{2 r^3}-\frac{56 a^2}{r^6}
+ \mathcal{O}\left(\frac{a^2b_0}{r^7}, \frac{a^4}{r^{10}}\right) , 
\label{K}
\end{align}
where 
% $\gamma$ denotes $\det(\gamma_{ij})$
%$\gamma^{(2)}_{ij}$ denotes the two-dimensional metric 
%in the equatorial plane $\theta = \pi/2$,
%and 
$a$ and $b_0$ are book-keeping parameters 
in the weak field approximation. 
As for the first line of Eq. (\ref{K}), please see e.g. 
the page 263 in Reference \cite{LL}.  
We note that the first term in the second line of Eq. (\ref{K}) 
does not contribute because $\Gamma^{\phi}_{rr} = 0$. 
It is not surprising that this Gaussian curvature does not agree with 
Eq. (26) in Jusufi and \"Ovg\"un \cite{JO}, 
because their Gaussian curvature describes 
another surface that is associated with the Randers-Finsler metric 
different from our optical metric. 

In order to perform the surface integral of the Gaussian curvature 
in Eq. (\ref{GB-axial}), 
we have to determine the boundary 
of the integration domain. 
In other words, we need the light ray as a function of $r(\phi)$. 
For the later convenience, we introduce the inverse of $r$ as 
$u \equiv r^{-1}$. 
The orbit equation in this case becomes 
\begin{align}
\left(\frac{du}{d\phi}\right)^2 
=&\frac{1}{b^2}-u^2-\frac{b_0u}{b^2}+b_0u^3-\frac{4au}{b^3}
-\frac{4a(b_0u-{b_0}^2u^2)}{b^3}
+\mathcal{O}\left(\frac{a^2}{b^6}\right), 
\end{align}
where $b$ is the impact parameter of the photon. 
See e.g. Reference \cite{OIA2017} on how to obtain the photon 
orbit equation in the axisymmetric and stationary spacetime. 
The orbit equation is iteratively solved as 
\begin{align}
u=\frac{\sin\phi}{b}+\frac{\cos^2\phi}{2b^2}b_0-\frac{2}{b^3}a
+\mathcal{O}\left(\frac{{b_0}^2}{b^3}, \frac{ab_0}{b^4}\right),
\label{u}
\end{align}

By using Eq. (\ref{u}) as the iterative solution for the photon orbit, 
the surface integral of the Gaussian curvature in Eq. (\ref{GB-axial}) 
is calculated as 
\begin{align}
-\iint_{{}^{\infty}_{R}\square^{\infty}_{S}} K dS 
=&\int_{\infty}^{r(\phi)} dr
\int_{\phi_S}^{\phi_R} d\phi \left(-\frac{b_0}{2r^2}\right) 
+\mathcal{O}\left(\frac{{b_0}^2}{b^2}, \frac{ab_0}{b^3}\right)
\notag\\
=&\frac{b_0}{2}
\int_{0}^{\frac{\sin\phi}{b}+\frac{\cos^2\phi}{2b^2}b_0-\frac{2}{b^3}a} du
\int_{\phi_S}^{\phi_R} d\phi  
+\mathcal{O}\left(\frac{{b_0}^2}{b^2}, \frac{ab_0}{b^3}\right)
\notag\\
%=&\frac{b_0}{2}\int_{\phi_S}^{\phi_R}
%\Big[u\Big]^{\frac{\sin\phi}{b}+\frac{\cos^2\phi}{2b^2}b_0-\frac{2}{b^3}a}_{0} d\phi 
%+\mathcal{O}({b_0}^2,ab_0)
%\notag\\
=&\frac{b_0}{2}\int_{\phi_S}^{\phi_R}
\Big[\frac{\sin\phi}{b}\Big] d\phi 
+\mathcal{O}\left(\frac{{b_0}^2}{b^2}, \frac{ab_0}{b^3}\right)
\notag\\
=&\frac{b_0}{2}\Big[-\frac{\cos\phi}{b}\Big]_{\phi=\phi_S}^{\phi_R} 
+\mathcal{O}\left(\frac{{b_0}^2}{b^2}, \frac{ab_0}{b^3}\right)
\notag\\
=&\frac{b_0}{2b}
\left(\sqrt{1-b^2{u_R}^2}+\sqrt{1-b^2{u_S}^2}\right)
+\mathcal{O}\left(\frac{{b_0}^2}{b^2}, \frac{ab_0}{b^3}\right) ,
\label{intK}
\end{align}
where we used $\sin{\phi_R} = bu_R +\mathcal{O}(ab^{-2}, b_0b^{-1})$ 
and $\sin{\phi_S} = bu_S +\mathcal{O}(ab^{-2}, b_0b^{-1})$ by Eq. (\ref{u}) 
in the last line.

\subsection{Geodesic curvature}
The geodesic curvature provides an important contribution 
to our calculations of the light deflection, though 
it is not usually described in standard textbooks on 
the general relativity. 
Hence, we follow Reference \cite{OIA2017} to 
briefly explain the geodesic curvature here. 
The geodesic curvature can be defined in the vector form as 
(e.g. \cite{Math})
\begin{align}
\kappa_g \equiv \vec{T}^{\prime} \cdot \left(\vec{T} \times \vec{N}\right) , 
\label{kappag-vector}
\end{align}
where we assume a parameterized curve with a parameter, 
$\vec{T}$ is the unit tangent vector for the curve 
by reparameterizing the curve using its arc length, 
$\vec{T}^{\prime}$ is its derivative with respect to the parameter, 
and $\vec{N}$ is the unit normal vector for the surface. 
Eq. (\ref{kappag-vector}) can be rewritten in the tensor form as 
\begin{align}
\kappa_g = \epsilon_{ijk} N^i a^j e^k , 
\label{kappag-tensor}
\end{align}
where $\vec{T}$ and $\vec{T}^{\prime}$ correspond to 
$e^k$ and $a^j$, respectively. 
Here, the Levi-Civita tensor 
$\epsilon_{ijk}$ is defined by 
$\epsilon_{ijk} \equiv \sqrt{\gamma}\varepsilon_{ijk}$, 
where 
$\gamma \equiv \det{(\gamma_{ij})}$, 
and $\varepsilon_{ijk}$ is the Levi-Civita symbol 
($\varepsilon_{123} = 1$). 
In the present paper,  we use $\gamma_{ij}$ in the above definitions 
but not $g_{ij}$. 
Note that $a^i \neq 0$ in the three-dimensional optical metric 
by nonvanishing $g_{0i}$ \cite{OIA2017}, 
even though the light signal follows a geodesic 
in the four-dimensional spacetime. 
On the other hand, we notice that if we would have a geodesics 
in the optical metric then
$a^i = 0$ and thus $\kappa_g = 0$. 

As shown first in Reference \cite{OIA2017}, 
Eq. (\ref{kappag-tensor}) is rewritten as 
\begin{align}
\kappa_g = - \epsilon^{ijk} N_i \beta_{j|k} , 
\label{kappag-tensor2} 
\end{align}
where we use $\gamma_{ij}e^ie^j = 1$. 

Henceforth, we focus on the equatorial plane ($\theta = \pi/2$). 
Then, let us denote the unit normal vector as $N_p$. 
This vector is normal to the $\theta$-constant surface. 
Therefore, it satisfies 
$N_p \propto \nabla_p \theta = \delta^{\theta}_p$, 
where $\nabla_p$ is the covariant derivative associated with
$\gamma_{ij}$. 
Hence, $N_p$ is written in a form as 
$N_p = N_{\theta} \delta^{\theta}_p$. 
By noting that $N_p$ is a unit vector ($N_pN_q \gamma^{pq} = 1$), 
we obtain $N_{\theta} = \pm 1/\sqrt{\gamma^{\theta\theta}}$. 
Therefore, $N_p$ can be expressed as 
\begin{align}
N_p = \frac{1}{\sqrt{\gamma^{\theta\theta}}} \delta_p^{\theta} , 
\label{N}
\end{align}
where we choose the upward direction without loss of generality. 

For the equatorial case, one can show 
\begin{align}
\epsilon^{\theta p q} \beta_{q|p} 
&=-\frac{1}{\sqrt{\gamma}}\beta_{\phi,r} , 
\label{rot-beta}
\end{align}
where the comma denotes the partial derivative, 
we use $\epsilon^{\theta r \phi} = - 1/\sqrt{\gamma}$ 
and 
we note $\beta_{r,\phi} = 0$ owing to the axisymmetry. 
By using Eqs. (\ref{N}) and (\ref{rot-beta}), 
the geodesic curvature of the light ray with the generalized optical metric 
becomes 
\cite{OIA2017}
\begin{align}
\kappa_g=-\sqrt{\frac{1}{\gamma\gamma^{\theta\theta}}}\beta_{\phi,r} . 
\label{kappa_equ}
\end{align}
For Teo wormhole case, this is obtained as 
\begin{align}
\kappa_g=&-\frac{2a}{r^3}
%-\frac{ab_0}{r^4}
+\mathcal{O}\left(\frac{a^3}{r^7}, \frac{a^3b_0}{r^8}\right) . 
\end{align}

We examine the contribution from the geodesic curvature. 
This contribution is the path integral along the light ray 
(from the source to the receiver),  
which is computed as 
\begin{align}
\int_S^R\kappa_gd\ell
=&\int^{S}_{~R}\frac{2a}{r^3} d\ell  
+\mathcal{O}\left(\frac{{b_0}^2}{b^2}, \frac{ab_0}{b^3}\right)
\notag\\
=&\int^{\pi/2-\phi_S}_{~\pi/2-\phi_R} \frac{2a\cos\vartheta}{b^2}
d\vartheta 
+\mathcal{O}\left(\frac{{b_0}^2}{b^2}, \frac{ab_0}{b^3}\right)
\notag\\
=&\frac{2a}{b^2}
\left[\sin\left(\frac{\pi}{2}-\phi_S\right)
-\sin\left(\frac{\pi}{2}-\phi_R\right)\right] 
+\mathcal{O}\left(\frac{{b_0}^2}{b^2}, \frac{ab_0}{b^3}\right)
\notag\\
=&\frac{2a}{b^2}\left(\sqrt{1-b^2{u_S}^2}+\sqrt{1-b^2{u_R}^2}\right) 
+\mathcal{O}\left(\frac{{b_0}^2}{b^2}, \frac{ab_0}{b^3}\right) , 
\label{intkappa}
\end{align}
for the retrograde case of the photon orbit. 
In the last line, we used 
$\sin{\phi_R} = bu_R +\mathcal{O}(ab^{-2}, b_0b^{-1})$ 
and $\sin{\phi_S} = bu_S +\mathcal{O}(ab^{-2}, b_0b^{-1})$ 
by Eq. (\ref{u}). 
The above contribution becomes $4a/b^2$, as $r_R \to \infty$ and 
$r_S \to \infty$. 
The sign of the right hand side of Eq. (\ref{intkappa}) changes, 
if the photon orbit is prograde.

\subsection{Deflection angle}
By combining Eqs. (\ref{intK}) and (\ref{intkappa}), 
the deflection angle of light for the prograde case 
is obtained as 
\begin{align}
\alpha_{\mbox{prog}}
=\frac{b_0}{2b}
\left(\sqrt{1-b^2{u_R}^2}+\sqrt{1-b^2{u_S}^2}\right)
-\frac{2a}{b^2}
\left(\sqrt{1-b^2{u_S}^2}+\sqrt{1-b^2{u_R}^2}\right) 
+\mathcal{O}\left(\frac{{b_0}^2}{b^2}, \frac{ab_0}{b^3}\right) . 
\label{alpha-prog}
\end{align}
The deflection angle for the retrograde case is 
\begin{align}
\alpha_{\mbox{retro}}=
\frac{b_0}{2b}
\left(\sqrt{1-b^2{u_R}^2}+\sqrt{1-b^2{u_S}^2}\right)
+\frac{2a}{b^2}
\left(\sqrt{1-b^2{u_S}^2}+\sqrt{1-b^2{u_R}^2}\right) 
+\mathcal{O}\left(\frac{{b_0}^2}{b^2}, \frac{ab_0}{b^3}\right) . 
\label{alpha-retro}
\end{align}
For both cases, the source and receiver may be located 
at finite distance from the wormhole. 
Eqs. (\ref{alpha-prog}) and (\ref{alpha-retro}) show that 
the light deflection is increasing with decreasing impact parameter 
and increasing throat radius.
%This can be understood by the following intuitive explanations. 
The light deflection in the prograde (retrograde) direction 
is decreasing (increasing) 
with increasing the angular momentum of the Teo wormhole, 
because the local inertial frame (in which the light propagates 
at the light speed $c$ in general relativity) 
moves faster (slower) and hence the light signal feels 
the gravitational pull for shorter (longer) time. 
Regarding the light propagation around a rotating object, 
similar physical explanations based on the dragging 
of the inertial frame were done about the Shapiro time delay 
by Laguna and Wolsczan 
\cite{LW}.

One can see that, in the limit as $r_R \to \infty$ and $r_S \to \infty$, 
Eqs. (\ref{alpha-prog}) and (\ref{alpha-retro}) become 
\begin{align}
\alpha_{\mbox{prog}} \rightarrow \frac{b_0}{b}-\frac{4a}{b^2} 
+\mathcal{O}\left(\frac{{b_0}^2}{b^2}, \frac{ab_0}{b^3}\right) ,
\notag\\
\alpha_{\mbox{retro}} \rightarrow \frac{b_0}{b}+\frac{4a}{b^2} 
+\mathcal{O}\left(\frac{{b_0}^2}{b^2}, \frac{ab_0}{b^3}\right) .
\end{align}
They agree with Eqs. (39) and (56) in Jusufi and \"Ovg\"un \cite{JO}, 
in which they are restricted within the asymptotic source and receiver 
($r_R \to \infty$ and $r_S \to \infty$).

\section{Conclusion}
In the weak field approximation, 
we have discussed the deflection angle of light for an observer and source 
at finite distance from a rotating Teo wormhole. 
We have shown that both of 
the Werner's method and the generalized optical metric method 
give the same deflection angle at the leading order of the weak field approximation, 
if the receiver and source are at the null infinity. 
We have also found corrections for the deflection angle 
due to the finite distance from the wormhole. 
It is left for future to study higher order terms in the weak field approximation 
of a rotating Teo wormhole and to examine also the strong deflection limit.

\begin{acknowledgments}
We are grateful to Marcus Werner for the useful discussions. 
We wish to thank Kimet Jusufi for his comments for 
clarifying his paper Ref. \cite{JO}. 
We would like to thank 
Yuuiti Sendouda, Ryuichi Takahashi, Yuya Nakamura and 
Naoki Tsukamoto 
for the useful conversations. 
This work was supported 
in part by Japan Society for the Promotion of Science (JSPS) 
Grant-in-Aid for Scientific Research, 
No. 17K05431 (H.A.),  No. 18J14865 (T.O.), 
in part by Ministry of Education, Culture, Sports, Science, and Technology,  
No. 17H06359 (H.A.)
and 
in part by JSPS research fellowship for young researchers (T.O.).  
\end{acknowledgments}


\begin{thebibliography}{99}
\bibitem{ER}
A. Einstein and N. Rosen, 
Phys. Rev. {\bf 48}, 73 (1935).
\bibitem{Wheeler}
J. A. Wheeler, Phys. Rev. {\bf 97}, 511 (1955); 
R. W. Fuller and J. A. Wheeler, Phys. Rev. {\bf 128}, 919 (1962). 
\bibitem{MW}
C. W. Misner and J. A. Wheeler, 
Ann. Phys. {\bf 2}, 525 (1957). 
\bibitem{MTY}
M. S. Morris and K. S. Thorne, 
Am. J. Phys. {\bf 56}, 395 (1988). 
M. S. Morris, K. S. Thorne and U. Yurtsever, 
Phys. Rev. Lett. {\bf 61}, 1446 (1988). 
\bibitem{Visser}
M. Visser, Phys. Rev. D {\bf 39}, 3182(R) (1989). 
\bibitem{Visser-Book}
M. Visser, {\it Lorentzian Wormholes: From Einstein to Hawking}, 
(NY, American Institute of Physics, 1995). 
\bibitem{GBWH}
E. Gravanis and S. Willison, 
Phys. Rev. D {\bf 75}, 084025 (2007). 
\bibitem{CC}
L. Chetouani, and G. Cl\'ement, 
Gen. Relativ. Gravit. {\bf 16}, 111 (1984). 
\bibitem{Clement}
G. Cl\'ement, 
Int. J. Theor. Phys. 23, 335 (1984). 
\bibitem{Perlick}
V. Perlick, 
Phys. Rev. D {\bf 69}, 064017 (2004). 
\bibitem{Safonova}
M. Safonova, D. F. Torres, and G. E. Romero, 
Phys. Rev. D {\bf 65}, 023001 (2001). 
\bibitem{Shatskii}
A. A. Shatskii, Astron. Rep. {\bf 48}, 525 (2004). 
\bibitem{Nandi}
K. K. Nandi, Y. Z. Zhang, and A. V. Zakharov, 
Phys. Rev. D {\bf 74}, 024020 (2006). 
\bibitem{BP}
A. Bhattacharya, and A. A. Potapov, 
Mod. Phys. Lett. A, {\bf 25}, 2399 (2010). 
\bibitem{Abe}
F. Abe, Astrophys. J. {\bf 725}, 787 (2010). 
\bibitem{Toki}
Y. Toki, T. Kitamura, H. Asada, and F. Abe, 
Astrophys. J. {\bf 740}, 121 (2011). 
\bibitem{Tsukamoto}
N. Tsukamoto, T. Harada, 
Phys. Rev. D {\bf 87}, 024024 (2013). 
\bibitem{Kitamura}
T. Kitamura, K. Nakajima, and H. Asada, 
Phys. Rev. D {\bf 87}, 027501 (2013). 
\bibitem{Sakai}
T. Ohgami and N. Sakai, 
Phys.Rev. D {\bf 91}, 124020 (2015); 
T. Ohgami and N. Sakai, 
Phys.Rev. D {\bf 94}, 064071 (2016); 
M. Kuniyasu, K. Nanri, N. Sakai, T. Ohgami, R. Fukushige, S. Komura
Phys. Rev. D {\bf 97}, 104063 (2018). 
\bibitem{JOB}
K. Jusufi, A. \"Ovg\"un and A. Banerjee
Phys. Rev. D {\bf 96}, 084036 (2017). 
\bibitem{Asada}
H. Asada, 
Mod. Phys. Lett. A, {\bf 32}, 1730031 (2017). 
\bibitem{DS}
T. K. Dey, and S. Sen, 
Mod. Phys. Lett. A, {\bf 23}, 953 (2008). 
\bibitem{Nakajima}
K. Nakajima, and H. Asada, Phys. Rev. D {\bf 85}, 107501 (2012).
\bibitem{Takahashi}
R. Takahashi, H. Asada, 
Astrophys. J. {\bf 768}, L16 (2013). 
\bibitem{JO}
Kimet Jusufi, and Ali \"Ovg\"un, 
Phys. Rev. D {\bf 97}, 024042, (2018). 
\bibitem{Teo}
E. Teo, 
Phys. Rev. D {\bf 58}, 024014 (1998).
\bibitem{GW}
G. W. Gibbons, M. C. Werner, 
Class. Quant. Grav. {\bf 25}, 235009 (2008). 
\bibitem{Werner2012}
M. C. Werner, 
Gen. Rel. Grav. {\bf 44}, 3047 (2012). 
\bibitem{OIA2017}
T. Ono, A. Ishihara, and H. Asada, 
Phys. Rev. D {\bf 96}, 104037 (2017). 
\bibitem{Abramowicz}
M. A. Abramowicz, B. Carter and J. P. Lasota, 
Gen. Relat. Grav., {\bf 20}, 1173 (1988).
\bibitem{GP}
G. W. Gibbons, M. J. Perry, 
Proc. R. Soc. London, Ser. A {\bf 358}, 467 (1978).
\bibitem{BH}
S. E. Perez Bergliaffa, K. E. Hibberd, 
ArXiv:gr-qc/0006041. 
\bibitem{Lewis}
T. Lewis, Proc. Roy. Soc. A, {\bf 136}, 176 (1932). 
\bibitem{LR}
H. Levy, and W. J. Robinson, 
Proc. Camb. Phil. Soc. {\bf 60}, 279 (1963). 
\bibitem{Papapetrou}
A. Papapetrou, 
Ann. Inst. H. Poincare A, {\bf 4}, 83 (1966). 
\bibitem{GB-theorem}
M. P. Do Carmo, 
{\it Differential Geometry of Curves and Surfaces}, 
pages 268-269, 
(Prentice-Hall, New Jersey, 1976).
\bibitem{Math}
A. C. Belton, 
{\it Geometry of Curves and Surfaces}, page 38 (2015); 
www.maths.lancs.ac.uk/~belton/www/notes/geom{}\underline{ }notes.pdf; 
J. Oprea, 
{\it Differential Geometry and Its Applications (2nd Edition)},  
page 210, 
(Prentice Hall, New Jersey, 2003). 
\bibitem{Ishihara}
A. Ishihara, Y. Suzuki, T. Ono, T. Kitamura, and H. Asada, 
Phys. Rev. D {\bf 94}, 084015 (2016); 
A. Ishihara, Y. Suzuki, T. Ono, and H. Asada, 
Phys. Rev. D {\bf 95}, 044017 (2017).
\bibitem{AK}
H. Asada, and M. Kasai, 
Prog. Theor. Phys. {\bf 104}, 95 (2000). 
\bibitem{LL}
L. D. Landau, E. M. Lifschitz, 
{\it The Classical Theory of Fields (Third English Edition)}, 
(Pegamon Press, Oxford, 1971).
\bibitem{LW}
P. Laguna, A. Wolszczan, 
Astrophys. J., {\bf 486}, L27 (1997).
\end{thebibliography}
\end{document}